\documentclass[prb,twocolumn,showpacs,letterpaper,floatfix,superscriptaddress]{revtex4}
\usepackage{graphicx}
\usepackage{amsmath}
\usepackage{amssymb}
\def\myref#1{(\ref{#1})}
\def\be{\begin{equation}}
\def\ee{\end{equation}}
\begin{document}

\title{Coherent resistance of a disordered 1D wire:
Expressions for all moments and evidence for non-Gaussian 
distribution}

\author{Pavel \surname{Vagner}}
\email{p.vagner@fz-juelich.de} 
\affiliation{Institut f\"{u}r Schichten und Grenzfl\"{a}chen,% 
Forschungszentrum J\"{u}lich GmbH, 52425 J\"{u}lich, Germany} 
\affiliation{Institute of Electrical Engineering, Slovak Academy% 
of Sciences, D\'{u}bravsk\'{a} cesta 9, 842 39 Bratislava,% 
Slovakia}

\author{Peter \surname{Marko\v{s}}}
\affiliation{Institute of Physics, Slovak Academy of Sciences,% 
D\'{u}bravsk\'{a} cesta 9, 842 28 Bratislava, Slovakia}

\author{Martin \surname{Mo\v{s}ko}}
\affiliation{Institute of Electrical Engineering, Slovak Academy%  
of Sciences, D\'{u}bravsk\'{a} cesta 9, 842 39 Bratislava,%  
Slovakia} 
\affiliation{Institut f\"{u}r Schichten und Grenzfl\"{a}chen,%  
Forschungszentrum J\"{u}lich GmbH, 52425 J\"{u}lich, Germany}

\author{Thomas \surname{Sch\"{a}pers}}
\affiliation{Institut f\"{u}r Schichten und Grenzfl\"{a}chen,% 
Forschungszentrum J\"{u}lich GmbH, 52425 J\"{u}lich, Germany}

\date{\today}

\begin{abstract}
We study coherent electron transport in a one-dimensional wire 
with disorder modeled as a chain of randomly positioned 
scatterers. We derive analytical expressions for all statistical 
moments of the wire resistance $\rho$. By means of these 
expressions we show analytically that the distribution $P(f)$ of 
the variable $f=\ln(1+\rho)$ is not exactly Gaussian even in the 
limit of weak disorder. In a strict mathematical sense, this 
conclusion is found to hold not only for the distribution tails 
but also for the bulk of the distribution $P(f)$.
\end{abstract}

\pacs{73.23.-b}
%\keywords{}
\maketitle

%%%%%%%%%%%%%%%%%%%%%%%%%%%%%%%%%%%%%%%%%%%%%%%
%   Body of the text
%%%%%%%%%%%%%%%%%%%%%%%%%%%%%%%%%%%%%%%%%%%%%%%
\section{Introduction}

It is known that a coherent electron wave in a disordered 
one-dimensional (1D) wire of infinite length is exponentially 
localized by an arbitrary weak disorder. 
\cite{Mott-61,Borland-61,Mott-90} The resistance $\rho$ of the 1D 
wire of length $L$ should therefore increase with $L$ 
exponentially. In fact, the resistance wildly fluctuates from wire 
to wire in an ensemble of macroscopically identical wires (with 
disorder in each wire being microscopically different) and what 
increases exponentially is the mean resistance and also the 
``typical'' resistance. \cite{Landauer-70,Anderson-80}

It has also become clear that the resistance $\rho$ is not a 
self-averaged quantity. \cite{Anderson-80} In fact, the resistance 
fluctuations are so huge that (i) the resistance dispersion 
exceeds the mean resistance many orders of magnitude, (ii) the 
higher moments of the resistance exceed the mean resistance even 
more drastically, and (iii) the mean resistance is much larger 
than the typical one. These features are due to the fact that the 
moments of $\rho$ are governed by extremely high resistances 
occurring with an extremely low (but nonzero) probability.

To avoid the absence of self-averaging, the distribution $P(f)$ of 
the variable $f=\ln(1+\rho)$ was studied instead of the 
distribution 
$P(\rho)$.\cite{Anderson-80,stare,Melnikov,Shapiro-87} In contrast 
to $P(\rho)$, distribution $P(f)$ is well localized around the 
mean value $\overline{f}$. It is commonly accepted that for long 
enough wires the bulk of the distribution $P(f)$ is described by 
the Gauss function
\begin{equation} \label{e11}
P(f) = \frac{1}{\sqrt{2\pi\Delta^2}}
\exp\left[-\frac{{(f-\bar{f})}^2}{2\Delta^2}\right]  ,  
\end{equation}
where $\Delta^2 \equiv \overline{f^2}-{\overline{f}}^2$ is the 
variance, while the tails of the distribution $P(f)$ are allowed 
to be non-universal and depend on the model of disorder. In the 
limit of weak disorder it is accepted that 
$\Delta^2=2\overline{f}$, i.e., that the distribution (\ref{e11}) 
obeys the single parameter scaling. The two-parameter scaling is 
accepted to appear for strong disorder, where $\Delta^2$ is not an 
unambiguous function of $\overline{f}$. \cite{crs} Interesting to 
note, the authors of Ref.~\onlinecite{deych} found two-parameter 
scaling also for weak disorder, namely for the Anderson 1D 
disorder at certain conditions.

In this paper we study coherent transport in a 1D wire with 
disorder modeled as a chain of randomly positioned scatterers. We 
derive analytically all statistical moments of the wire 
resistance. By means of these moments we prove in the limit of 
long wires, that the distribution $P(f)$ always deviates from the 
Gauss distribution. The form of $P(f)$ for $f>\overline{f}$ is 
concluded to be nonuniversal (dependent on the model of disorder) 
even in the limit of weak disorder. In other words, in realistic 
wires disorder is never weak enough for $P(f)$ to be exactly 
Gaussian. The only approximation of our analysis is the phase 
randomization hypothesis. We confirm its validity by numerical 
simulations.

In Sec.~II we specify two different model of disordered 1D wire. 
As a model I we consider the statistical ensemble of wires with 
the same number of scatterers in each wire, in the model II we let 
the number of scatterers to fluctuate from wire to wire. In 
Sec.~III the moments of the wire resistance are derived for both 
models analytically assuming the phase randomization hypothesis. 
This hypothesis is verified in Sect.~IV by means of numerical 
simulations. In Sec.~V we prove that our expressions for the 
resistance moments are not consistent with the Gaussian form of 
$P(f)$ even in the limit of weak disorder. Discussion is given in 
Sect.~VI.

%%%%%%%%%%%%%%%%%%%%%%%%%%%%%%%%%%%%%%%%%%%%%%%
\section{Model of disordered 1D wire}

We consider a 1D wire with disorder represented by random 
potential

\noindent
\be\label{v}
V(x)=\sum _{i=1}^N \gamma\delta(x-x_i),
\ee
where $\gamma\delta(x-x_i)$ is the $\delta$-shaped impurity 
potential of strength $\gamma$, $x_i$ is the $i$-th impurity 
position selected at random along the wire, and $N$ is the number 
of impurities in the wire. Since the positions $x_i$ are mutually 
independent, the distances $a=x_{i+1}-x_i$ between the neighboring 
impurities follow the distribution $P(a)=N_I\exp[-N_Ia]$, where 
$N_I$ is the 1D density of impurities and $N_I^{-1}$ is the mean 
distance between the neighboring impurities.

In the following sections we examine two models. In the model I we 
consider the statistical ensemble of wires with $N$ fixed in each 
wire to its mean value $\langle N\rangle$. In the model II we fix 
the wire length $L$ and we let $N$ to fluctuate from wire to wire 
according to the distribution
\begin{equation} \label{e13}
{\mathcal G}(N)=\langle N\rangle^Ne^{-\langle N\rangle}/N! .
\end{equation}
It is easy to show that this distribution follows from the 
distribution $P(a)$. In both models $\langle N\rangle \equiv 
LN_I$.

The wire resistance $\rho$ (in units $h/2e^2$) is given by the 
Landauer formula \cite{Landauer-70}
\begin{equation} \label{e1}
\rho=\frac{R(\varepsilon_F)}{T(\varepsilon_F)},
\end{equation}
where $R$ and $T$ are the reflection and transmission coefficients 
describing the electron tunneling through disorder at the Fermi 
energy.

Using eq.~(\ref{e1}) we follow a number of previous localization 
practitioners. Instead of eq.~(\ref{e1}) we could use the 
two-terminal resistance $\rho=1/T=R/T+1$, which involves an extra 
term (unity on the right hand side) representing the fundamental 
resistance of contacts. The resistance~(\ref{e1}) thus represents 
the resistance of disorder, directly measurable only by four-probe 
techniques. The problem is that eq.~(\ref{e1}) ignores the effect 
of measurement probes \cite{Landauer-90,Datta-95}. We wish to note 
that this is not a serious problem in our case. First, we examine 
the regime $R/T \gg 1$, for which the two-terminal resistance 
$\rho=R/T+1$ coincides with eq.~(\ref{e1}). Second, with 
$\rho=R/T+1$ we would arrive at the same conclusions as with 
eq.~(\ref{e1}). Third, in principle, one can measure $R/T$ 
indirectly, by measuring the two-terminal resistance and then 
subtracting unity.

For disorder~(\ref{v}) both $R$ and $T$ can be obtained by solving 
the tunneling problem
\begin{equation} \label{e3}
\left[-\frac{\hbar^2}{2m}\
\frac{d^2}{dx^2}+V(x)\right] \Psi_k(x)=\mathcal{E}\Psi_k(x)
\end{equation}
with boundary conditions
\begin{equation} \label{e4a}
\Psi_k\left(x\rightarrow0\right)=e^{ikx}+r_k e^{-ikx},~~~
\Psi_k\left(x\rightarrow L\right)=t_k e^{ikx},
\end{equation}
where $\mathcal{E}=\hbar^2k^2/2m$ is the electron energy, $m$ is 
the effective mass, and $r_k$ and $t_k$ are the reflection and 
transmission amplitudes. The coefficients $R={|r_k|}^2$ and 
$T={|t_k|}^2$ need to be evaluated at the Fermi wave vector $k = 
k_F$.

The reflection coefficient of a single $\delta$-barrier is given 
as $R_I= \Omega^2/(k_F^2+\Omega^2)$, where 
$\Omega=m\gamma/\hbar^2$. We fix
\be\label{kfermi}
k_F=7.9\times 10^7 {\text{m}}^{-1}
\ee
and $m=0.067m_0$, and we parameterize the $\delta$-barrier by 
$R_I$. We ignore the fluctuations of $R_I$ as well as the spread 
of the impurity potentials.

%%%%%%%%%%%%%%%%%%%%%%%%%%%%%%%%%%%%%%%%%%%%%%%
\section{Resistance moments}
%%%%%%%%%%%%%%%%%%%%%%%%%%%%%%%%%%%%%%%%%%%%%%%
\subsection{Model I}

We start with derivation of the mean resistance. Assume that we 
know the reflection coefficient $R_N$ of a specific configuration 
of $N$ randomly positioned impurities. If we add to this 
configuration an extra impurity at position $x_{N+1}$, we can 
express $R_{N+1}$ through $R_N$ and $R_I$. It is useful to express 
$R_{N+1}$ in the form \cite{Landauer-70}
\begin{equation} \label{AppRho12}
\frac{R_{N+1}}{1-R_{N+1}} =
\frac{R_N + R_I - 2 \sqrt{R_N R_I} \cos \phi_N} {(1-R_N)
(1-R_I)} ,
\end{equation}
where $\phi_N$ is the phase specified below. Writing
eq.~(\ref{AppRho12}) in terms of the wire resistance
\be\label{xy}
\rho_N \equiv \frac{R_N}{1-R_N}
\ee
and in terms of
\be\label{l1}
\lambda_1 \equiv \frac{1+R_I}{1-R_I}
\ee
we get
\begin{equation} \label{rhoInak}
\rho_{N+1}= \lambda_1 \rho_N +
\frac{\lambda_1-1}{2} - \sqrt{(\lambda_1^2 -1) (\rho_N+\rho^2_N)}
\cos \phi_N .
\end{equation}
The phase $\phi_N = 2k_{F}a + \phi_0$, where $a=x_{N+1}-x_N$ is 
the inter-impurity distance, and $\phi_0$ is the ($a$-independent) 
phase due to the reflection by the obstacles. 
\cite{Landauer-70,Anderson-80} Obviously,
\be\label{ic}
\rho_0\equiv 0
\ee
and
\be\label{ic1}
\rho_1 \equiv \frac{R_I}{1-R_I} = \frac{\lambda_1-1}{2} .
\ee
Note that $\rho_2$ depends on $\phi_1$, $\rho_3$ depends on $\phi_2$
and $\phi_1$, etc., $\rho_{N+1}$ thus depends on $\phi_{N}$,
$\phi_{N-1}$, $\dots$, $\phi_{2}$, and $\phi_{1}$.

If we assume that $a \gg 2\pi/k_F$, then $\phi_N$ changes rapidly 
with $a$ and fluctuates at random from sample to sample as $a$ 
fluctuates. The ensemble average of $\rho_{N+1}$ over the 
inter-impurity distance $x_{N+1}-x_N$ then simplifies to 
\cite{Landauer-70,Anderson-80}
\begin{equation} \label{Apprho}
\overline {\rho_{N+1}} = \frac{1}{2\pi} \, \int \limits_0 ^{2\pi} 
d\phi_N \, \rho_{N+1} .
\end{equation}
If we average eq.~(\ref{rhoInak}) over $\phi_N$, the term $\propto 
\cos \phi_N$ becomes zero. If we then average over $\phi_{N-1}$, 
\dots, $\phi_2$, $\phi_1$, we obtain the recursion equation
\be \label{AppwN}
\overline{\rho_{N+1}} = \lambda_1\overline{\rho_{N}} + 
\frac{1}{2}(\lambda_1-1) .
\ee
We solve Eq. (\ref{AppwN}) with initial condition~(\ref{ic}) and 
obtain the mean resistance
\be\label{r1}
\overline{\rho_N}=\frac{1}{2}\left(\lambda_1^N-1\right).
\ee

The higher moments can be obtained in the same way. The $m$th 
power of eq.~(\ref{rhoInak}) averaged over $\phi_N$ formally reads
\begin{equation} \label{rmInak}
\overline{\rho^m_{N+1}} = \overline{\left[\lambda_1 \rho_N +
\frac{\lambda_1-1}{2} - \sqrt{(\lambda_1^2 -1) (\rho_N+\rho^2_N)}
\cos \phi_N \right]^m} .
\end{equation}
If we take into account that~\cite{AS}
\begin{equation} \label{CosInt}
\frac{1}{2\pi} \int \limits _0 ^{2\pi} d\phi \left( \cos \phi
\right)^{2m} = \frac{1 } {2^{2m}} \binom{2m}{m}
\end{equation}
and
\begin{equation} \label{CosInt0}
\frac{1}{2\pi} \int \limits _0 ^{2\pi} d\phi \left( \cos \phi
\right)^{2m-1} = 0 ,
\end{equation}
we easy see that eq.~(\ref{rmInak}) takes the form
\be\label{xx1}
\overline{\rho_{N+1}^m}=\sum_{k=0}^m \alpha_k(m) \rho_N^k ,
\ee
where coefficients $\alpha_k(m)$ are polynomial functions of
$\lambda_1$. Averaging each $\rho_N^k$ over $\phi_{N-1}$, each
$\rho_{N-1}^k$ over $\phi_{N-2}$, etc., we finally obtain the
recursion relation
\be\label{xx}
\overline{\rho_{N+1}^m}=\sum_{k=0}^m \alpha_k(m) \overline{\rho_N^k} .
\ee
A general expression for coefficients $\alpha_k(m)$ is given in the
Appendix~\ref{AlphApp}, where we also derive
\be\label{alphm}
\alpha_m(m) = \overline{ \left[ \lambda_1 - \sqrt{\lambda_1^2-1} 
\cos \phi \right]^m } .
\ee
We can also obtain eq.~(\ref{alphm}) by comparing the right hand 
sides of Eqs.~(\ref{rmInak}) and~(\ref{xx1}) for $\rho_N \to 
\infty$, where they reduce to $\overline{[\lambda_1 -
\sqrt{\lambda_1^2-1}\cos\phi]^m} \rho_N^m$ and $\alpha_m(m) 
\rho_N^m$, respectively.

We solve Eq.~(\ref{xx}) recursively. Suppose that the $N$-dependence
of $\overline{\rho^m_N}$ can be expressed in the form
\be\label{gf}
\overline{\rho_N^m}=a_m(m)\lambda_m^N+\dots a_1(m)\lambda_1^N+a_0(m).
\ee
For $m=1$ eq.~(\ref{gf}) coincides with eq.~(\ref{r1}). Therefore,
$\lambda_1$ in eq.~(\ref{gf}) coincides with eq.~(\ref{l1}) and
$a_1(1)=1/2$, $a_0(1)=-1/2$. Once we know $\lambda_1$, $a_1(1)$, and
$a_0(1)$, we can solve the problem for $m=2$ and determine
$\lambda_2$, $a_2(2)$, $a_1(2)$, and $a_0(2)$ (see the Appendix
\ref{a2App}). Generally, once we determine all $\lambda_k$ and all
coefficients $a_n(k)$ for $0\le n\le k\le m-1$, we can insert
expansion~(\ref{gf}) into eq.~(\ref{xx}) and compare the
$N$-independent factors at all $\lambda_{k\le m}^N$. This gives us
linear equations
\begin{equation} \label{aimFinal}
a_k(m) \lambda_k = \sum \limits _{i=k} ^{m} \alpha_i(m) \, a_k(i)
\end{equation}
for all $a_k(m)$ with $k<m$ and in addition the identity
$\lambda_m \equiv \alpha_m(m)$, i.e.,
\be\label{lm}
\lambda_m = \overline{ \left[ \lambda_1 - \sqrt{\lambda_1^2-1} 
\cos \phi \right]^m} .
\ee
As a last step we calculate the coefficient $a_m(m)$ with help of
the initial condition~(\ref{ic}). In the Appendix \ref{a2App} this
procedure is demonstrated in detail for $m = 2$. The result is
\be\label{r2a}
\overline{\rho_N^2} = \frac{1}{6} \lambda_2^N - \frac{1}{2} 
\lambda_1^N + \frac{1}{3} ,
\ee
where
\be\label{l2}
\lambda_2=\frac{1}{2}  \left(3\lambda_1^2 -1 \right).
\ee
Parameters $\lambda_k$ characterize the exponential increase of 
$\overline{\rho^m_N}$ with $N$. Equation~(\ref{lm}) expresses 
$\lambda_k$ analytically for arbitrary $k$, for example, for $m = 
1$ and $2$ it reproduces relations~(\ref{l1}) and~(\ref{l2}), 
respectively. We present also
\begin{eqnarray} \label{LamAll}
\lambda_3&=& \frac {5}{2} \, \lambda_1^{3} - \frac {3}{2} \, 
\lambda_1 ,
\\
\lambda_4&=& \frac {35}{8} \, \lambda_1^{4} - \frac {15}{4} 
\, \lambda_1^{2} + \frac {3}{8} ,
\\
\lambda_5&=&\frac {63}{8} \, \lambda_1^{5} - \frac {35}{4} \, 
\lambda_1^{3} + \frac {15}{8} \, \lambda_1 ,
\\
\lambda_6&=&\frac {231}{16} \, \lambda_1^{6} -
\frac {315}{16} \, \lambda_1^{4} +
\frac {105}{16} \, \lambda_1^{2} - \frac {5}{16} .
\end{eqnarray}
We do not present explicitly complete expressions for moments
$\overline{\rho_N^m}$ higher than $\overline{\rho_N^2}$. For further
purposes we only express the leading term of $\overline{\rho_N^m}$.
We see from (\ref{lm}) that $\lambda_1<\lambda_2<\dots<\lambda_m$.
Therefore, for large enough $N$
\begin{equation} \label{rmAprx}
\overline{\rho^m _N } \approx a_m(m)\, \lambda_m^N \propto
\lambda_m^N .
\end{equation}

For completeness, we derive also the mean value of the variable 
$f$. As in Ref. \onlinecite{Anderson-80}, we average over all 
phases the variable $f_N= \ln \left( 1+\rho_N \right)$ and obtain 
the recursion relation $\overline{f_{N+1}} = - \ln ( 1-R_I) + 
\overline{f_{N}}$. We solve this equation with the condition 
$\rho_0\equiv 0$ (i.e, with $\overline f_0\equiv 0$) and obtain
\begin{equation} \label{rtyp}
\overline{f_N} =- N \ln(1-R_I).
\end{equation}
No simple analytic expressions exist for higher moments 
$\overline{f^m}$. For details see 
Refs.~\onlinecite{Anderson-80,stare,crs}.

%%%%%%%%%%%%%%%%%%%%%%%%%%%%%%%%%%%%%%%%%%%%%%%
\subsection{Model II}

In the preceding section the number of impurities, $N$, was kept 
at the same value for each wire in the wire ensemble (model I). In 
this section we let $N$ to fluctuate from wire to wire according 
to the distribution~(\ref{e13}) while keeping for each wire the 
same wire length $L$ (model II). Thus, to obtain the resistance 
moments for the model II we just need to average over the 
distribution~(\ref{e13}) the moments obtained in the preceding 
section. In particular,
\be\label{stredovanie}
\langle \lambda^N_m\rangle=\sum_{N=1}^\infty
\lambda_m^N \mathcal{G}_N= e^{(\lambda_m-1)N_IL}=e^{m(m+1)L/\xi_m}  ,
\ee
where we define the $m$th characteristic length
$\xi_m$ as
\be\label{xim}
\xi_m^{-1}= \frac {N_I\left(\lambda_m-1\right)}{m(m+1)}.
\ee
From Eqs.~(\ref{r1},\ref{stredovanie}) we obtain the mean resistance
\be\label{tra}
\overline{\rho}=\frac{1}{2}(e^{2L/\xi_1}-1)
\ee
and from Eqs.~(\ref{r2a},\ref{stredovanie}) the 2nd moment
\be\label{trb}
\overline{\rho^2} = \frac{1}{6} e^{6L/\xi_2} - 
\frac{1}{2}e^{2L/\xi_1}+\frac{1}{3} .
\ee
The typical resistance is defined as $\rho_t=\exp{\overline f}-1$. 
We average $\overline f$ (eq.~\ref{rtyp}) over the distribution 
(\ref{e13}) and obtain
\be\label{rt}
\rho_t=\exp{L/\xi}-1 ,
\end{equation}
where
\begin{equation} \label{XiTyp}
\xi^{-1} = -N_I \ln (1-R_I) = N_I \ln\left( \frac{\lambda_1 +1}{2}
\right).
\end{equation}
is the electron localization length. For comparison,
\be\label{tra1}
\xi_1^{-1}=N_I\frac{R_I}{1-R_I}=N_I\left( \frac{\lambda_1 -1}{2} 
\right).
\ee
It is easy to show~\cite{comments} that $L/\xi_m$ can be expressed 
as an unambiguous function of $L/\xi$ a $L/\xi_1$. This means that 
our models exhibit two-parameter scaling. Only if $R_I$ is very 
small, both lengths converge to the same limit
\begin{equation} \label{e14}
\xi\approx\xi_1\approx{(N_IR_I)}^{-1}\quad, \quad R_I\ll 1.
\end{equation}
However, $\xi\ne\xi_1$ for any nonzero $R_I$.

%%%%%%%%%%%%%%%%%%%%%%%%%%%%%%%%%%%%%%%%%%%%%%%
\section{Microscopic modeling}

Our derivation of resistance moments relies on the phase 
randomization hypothesis, i.e., on the averaging~(\ref{Apprho}). 
This should be justified in the limit $a \gg 2\pi/k_F$, that means 
for $1/N_I \gg 2\pi/k_F$. Now we test the phase randomization 
hypothesis by microscopic modeling.

\begin{figure}[t]
\centerline{\includegraphics[clip,width=8.2cm]{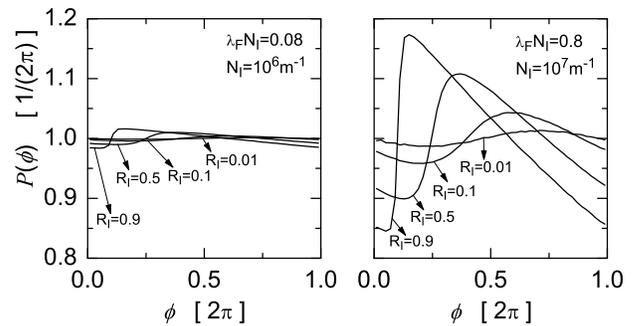}}
\caption{%
Distribution $P(\phi)$ for various model parameters.}%
\label{faza}
\end{figure}

In our microscopic model we select disorder as discussed in Sect. 
II, solve Eq.~(\ref{e3}) by the transfer matrix method, 
\cite{transfermatrix} and obtain from Eq.~(\ref{e1}) the 
resistance of a single wire. We repeat this process for a 
statistical ensemble of wires typically involving $10^6$ - $10^9$ 
samples.

In Fig.~\ref{faza} we present the distribution $P(\phi)$ of the 
variable $\phi$, where $\phi$ is the phase entering the right hand 
side of eq.~(\ref{rhoInak}). The distribution $P(\phi)$ can be 
accumulated either within the ensemble of wires with just two 
randomly positioned impurities within each wire, or within a 
single wire into which many impurities are positioned one by one. 
Both procedures give the same results.

In accord with the phase randomization hypothesis~(\ref{Apprho}), 
for low impurity density $N_I$ (left panel) we see that 
$P(\phi)\approx $ const = $1/(2\pi)$. Note that the flat 
distribution survives for rather large $R_I$ values. On the other 
hand, when $N_I$ is large, it tends to destroy the flatness of 
$P(\phi)$ even for very small values of $R_I$ (right panel).

\begin{figure}[t]
\centerline{\includegraphics[clip,width=8.2cm] {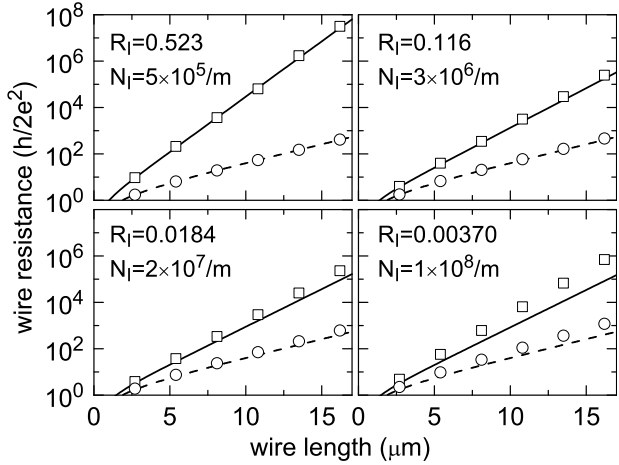}} 
\caption{%
Mean resistance (squares, full lines) and typical resistance 
(circles, dashed lines) versus the wire length $L$. Squares and 
circles are the microscopic model results, full lines and dashed 
lines are graphic representation of the formulas (\ref{tra}) and 
(\ref{rt}), respectively. $k_F$ is given in Eq. \ref{kfermi}. 
Parameters $N_I$ and $R_I$ are varied in such way that the 
localization length is the same ($\xi=2.7$~$\mu$m) for each panel. 
The accuracy of Eqs. (\ref{tra}) and (\ref{rt}) deteriorates with
increasing $N_I$.}%
\label{Fig7}
\end{figure}

Results presented in Fig.~\ref{faza} are consistent with those in 
Fig.~\ref{Fig7} where the mean and typical resistances obtained by 
microscopic modeling are presented for various reflection 
coefficients $R_I$ and various densities $N_I$. For low $N_I$ our 
microscopic data agree well with our analytical results. Note that 
this is the case also for large $R_I$. However, with increasing 
$N_I$ the agreement deteriorates.

%%%%%%%%%%%%%%%%%%%%%%%%%%%%%%%%%%%%%%%%%%%%%%%
\section{Moments of the resistance in the limit of very long wires.}

\begin{figure}[t]
\centerline{\includegraphics[clip,width=8.2cm]{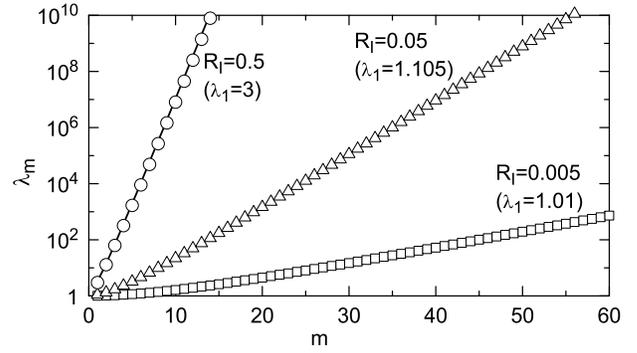}}
\caption{%
$m$-dependence of $\ln\lambda_m$ for various $R_I$.}%
\label{fig_3}
\end{figure}

In the limit of long wires $\langle N\rangle$ becomes large and 
only the leading term of the moment $\overline{\rho^m_N}$ becomes 
important. From Eqs.~(\ref{rmAprx}) and~(\ref{stredovanie}) one 
easy obtains
\be\label{rozdiel}
\overline{\rho^m}\propto\left\{
\begin{array}{ll}
\lambda_m^{\langle N\rangle}          &~~\text{Model~I}\\
~~~                      &   ~~~                  \\
e^{\langle N\rangle(\lambda_m-1)}  &~~\text{Model~II}
\end{array}\right.
\ee
\noindent
From Eq. \ref{lm} it is evident that

\begin{equation} \label{mLamL}
\ln\lambda_m\approx m~~~~~~~m\gg 1
\end{equation}
\noindent
In Fig.~\ref{fig_3} the estimate~(\ref{mLamL}) is verified numerically.
Using eq.~(\ref{mLamL}) and $\langle N\rangle \equiv LN_I$ we 
can obtain from~(\ref{rozdiel})
\be\label{gt}
\ln \overline{\rho^m(L)}\propto \left\{
\begin{array}{ll}
mL  & ~~\text{Model~I}\\
\\
L e^{m \eta}  &  ~~\text{Model~II}
\end{array}\right.
\ee

Now we show that the analytical formulae~(\ref{gt}) 
are not consistent with the assumption that 
the distribution $P(f)$ is Gaussian. To see this clearly, let 
us average the $m$th power of the resistance
\be\label{gf00}
\overline{\rho^m(L)}=\int_0^\infty df \ P(f)(e^f-1)^m
\ee
over the Gauss distribution~(\ref{e11}). The result can easy be 
obtained analytically as
\be \label{gf0}
\overline{\rho^m(L)}=\sum_{k=1}^m(-1)^{m-k}
\binom{m}{k} \exp\left(\frac{k^2\Delta^2}{2}+k\bar{f}\right)  .
\ee
In the limit $L/\xi\gg 1$ relation (\ref{gf0}) reduces to
\be\label{gf01}
\overline{\rho^m(L)} = \exp\left( \frac{m^2\Delta^2}{2}+ m 
\overline{f}\right).
\ee
Since $\Delta^2\propto L$, from~(\ref{gf01})we have
\be\label{m2}
\ln \overline{\rho^m(L)}\propto m^2L.
\ee
In particular, for weak disorder $\Delta^2=2\bar{f}=2L/\xi$ and 
the leading term in the sum~(\ref{gf0}) reads $\propto 
e^{m(m+1)L/\xi}$.

If we compare eq.~(\ref{m2}) with our analytical results 
(\ref{gt}), we immediately see that relations~(\ref{gt}) do not 
approach the dependence $\propto m^2L$ predicted by 
relation~(\ref{m2}). Since the higher moments of the resistance 
are mainly governed by the distribution $P(f)$ for 
$f>\overline{f}$, the difference between the relations~(\ref{m2}) 
and~(\ref{gt}) is a proof that $P(f)$ deviates from the Gauss 
distribution in the model I as well as in the model II.

It is important to note that these deviations are not restricted 
to the distribution tail $f\gg \overline{f}$. It is known that the 
tail of the distribution $P(f)$ is non-universal. From Eq. 
\myref{rhoInak} we see that $\rho_{N+1} \le (\lambda_1 + 
\sqrt{\lambda_1^2-1}) \rho_N$. Therefore, in the model I the value 
of $f$ never exceeds the maximum value $f_{\text{max}}$ given by
\be\label{fmax}
\frac {f_{\text{max}}} {\overline{f}} \approx 
\frac{ \ln \left( \lambda_1 + \sqrt{\lambda_1^2-1} \right)}
{\ln\left[(\lambda_1+1)/2\right]}.
\ee
Due to this reason, in the model I the distribution $P(f)$ drops 
to zero for $f>f_{\text{max}}$ and some deviations from the Gauss 
distribution \myref{e11} can be expected to appear already for $f$ 
slightly below $f_{\text{max}}$. The same holds also for the model 
II in which $N$ fluctuates so that the difference $f_{\text{max}} 
(\text{model II}) - f_{\text{max}}( \text{model I})$ is of order 
of $\sqrt{N}$. This means that the distribution $P(f)$ drops to 
zero in both models if $f$ is large enough.

However, this sudden drop to zero is not responsible for the 
nonGaussian behavior represented by eq.~\myref{gt}. To prove this 
we now show that $\overline{\rho^m}$ is governed by the $f$ values 
much smaller that $f_{\text{max}}$. We show that the maximum of 
the function $P(f)\times e^{mf}$ is positioned at $f=f_m$, where 
$f_m$ is much smaller than $f_{\text{max}}$. For the Gaussian 
distribution \myref{e11} we find
\begin{equation}
\frac{f_m}{\overline{f}}= 
\frac{\overline{f}+m\Delta^2}{\overline{f}} = (2m+1) .
\end{equation}
This ratio depends neither on $L$ nor on $\lambda_1$. Note that 
the ratio $f_{\text{max}}/\overline{f}$ does not depend on $L$ but 
it still depends on $\lambda_1$. In the limit of weak disorder 
($R_I\to 0$) we obtain $\lambda_1\approx 1+2R_I$ and 
$f_{\text{max}}/\overline{f} \sim 2 / \sqrt{R_I} \to 
\infty$. It is thus evident that $f_m \ll f_{\text{max}}$ at least 
in the limit of weak disorder.

From~(\ref{gt}) we obtain
\begin{equation}
\frac{ \ln \overline{\rho^{m+1}(L)} }
{ \ln \overline{ \rho^{m}(L)} } = \frac{m+1}{m}
\end{equation}
for the model I, while for the Gaussian distribution  
\begin{equation}
\frac{ \ln \overline{\rho^{m+1}(L)} }
{ \ln \overline{ \rho^{m}(L)} } = \frac{m+2}{m}
\end{equation}
This proves that $P(f)$ deviates the Gaussian distribution already 
for $f$ from the neighbourhood of $f_1$. As discussed above, this 
region is still far from the distribution tail.

In the model II this deviation from the Gaussian shape is even 
more pronounced, because $\overline{\rho^m(L)}$ increases with $m$ 
much faster than the dependence~(\ref{gf01}). This means that 
$P(f)$ decreases for $f>\overline{f}$ much slower than the 
Gaussian distribution. The slower decrease means that the 
deviation from Gaussian is surely not caused by the cutoff at 
$f=f_{\text{max}}$.

%%%%%%%%%%%%%%%%%%%%%%%%%%%%%%%%%%%%%%%%%%%%%%%
\section{Discussion and conclusions}

In conclusion, we have presented two simple models of disordered 
wire which allowed us to express analytically all moments of the 
wire resistance. By means of these analytical expressions we have 
succeeded to prove analytically the nonGaussian behavior of the 
distribution $P(f)$.

Analytical formulae for the resistance moments were obtained 
assuming the phase randomization hypothesis. In Sect. IV we have 
proven numerically that this hypothesis is indeed valid for small 
impurity density $N_I$. This means that for small enough $N_I$ our 
results are exact.

In fact, in a strict mathematical sense there is no single 
parameter scaling in the models I and II, because the lengths 
$\xi$ and $\xi_1$ always differ from each other. The difference 
between them is very small in the limit of small reflection 
coefficient, $R_I\ll 1$. Then, numerical experiment is not able do 
distinguish between $\xi$ and $\xi_1$ and the single parameter 
scaling holds to a good approximation for the bulk of the 
distribution.

\begin{figure}[t]
\centerline{\includegraphics[clip,width=8.2cm]{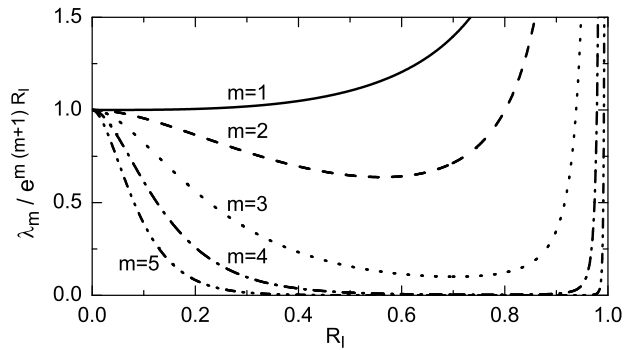}}
\caption{%
The ratio $\lambda_m/\exp\left[m(m+1)R_I\right]$ as a function of 
$R_I$ for $m=$1 -- 5.}%
\label{fig_4}
\end{figure}

If we accept $\xi=\xi_1$ as in eq.~(\ref{e14}), then the relations 
(\ref{tra}), (\ref{trb}) and (\ref{rt}) agree with those derived 
within the scaling theory of localization. \cite{stare} Note that 
the relation (\ref{e14}) is exact only if the second and higher 
orders of $R_I$ can be neglected. The same condition assures the 
equivalence of eq. (\ref{m2}) with eq. (\ref{gt}). Indeed, if we 
expand $\lambda_m$ [eq.~(\ref{lm})] into powers of $R_I$ and 
neglect all higher powers of $R_I$, we can interpret the obtained 
``expansion'' as the first two terms of the Taylor expansion of 
the exponential function, i.e.,
\be\label{taylor}
\lambda_m= 1+m(m+1)R_I + \mathcal{O}(R_I^2) \approx 
e^{\left[m(m+1)R_I\right]}.
\ee
We show in Fig. \ref{fig_4} that the approximation (\ref{taylor}) 
is very good in the limit of very small $R_I$ and small $m$. 
However, for any $R_I$ we can find such $m$ that the approximation 
(\ref{taylor}) is no longer valid. Therefore, relation (\ref{m2}) 
does not give the correct $R_I$ dependence for higher moments of 
resistance. This proves that the distribution $P(f)$ is not 
Gaussian even for an infinitesimally small $R_I$.

The main difference between the presented results and those of the 
scaling theory of localization is that in our model we keep the 
exact $R_I$ dependence of all $\lambda_m$s while in the scaling 
theory only the linear term in $R_I$ is kept. To understand this 
difference more clearly, let us go back to the relation 
(\ref{AppwN}). We can approximate $\lambda_1$ as in (\ref{taylor}) 
and rewrite (\ref{AppwN}) as
\be\label{A1}
\overline{\rho_{N+1}}=\delta\rho+\overline{\rho_N}+2\delta\rho~\overline{\rho_N}.
\ee
Equation (\ref{A1}) is formally identical with the recursion 
relation derived in Refs. \onlinecite{Anderson-80,Melnikov}. 
However, in these works it is supposed that the increment 
$\delta\rho$ is proportional to the increment $\delta L$ of the 
wire length. The terms of higher order in $\delta\rho$ can 
therefore be neglected and the approximation (\ref{taylor}) 
becomes exact. This is not the case in our model, where 
$\delta\rho=R_I$ does not depend on the length scale and it is not 
possible to perform the limit $\delta L\to 0$. As this limit plays 
a crucial role in the derivations of SPS in Refs. 
\onlinecite{Anderson-80} and \onlinecite{Melnikov}, it is 
understandable that our model does not provide us with the Gauss 
distribution of $f$ predicted by these derivations. \cite{pozna}

\begin{acknowledgments}
P.V. was supported by a Marie Curie Fellowship of the Fifth 
Framework Programme of the European Community, contract no. 
HPMFCT-2000-00702. M.M. and P.V. were also supported by the VEGA 
grant no. 2/7201/21. M.M., P.V., and P.M. were also supported by 
the Science and Technology Assistance Agency under Grant No. 
APVT-51-021602.
\end{acknowledgments}

%%%%%%%%%%%%%%%%%%%%%%%%%%%%%%%%%%%%%%%%%%%%%%%%%%%%%%%%%%%%%%%%%%%
% APPENDIX
%%%%%%%%%%%%%%%%%%%%%%%%%%%%%%%%%%%%%%%%%%%%%%%%%%%%%%%%%%%%%%%%%%%
\appendix
\section{} \label{AlphApp}

Coefficients $\alpha_k(m)$ in Eqs.~(\ref{xx1}, \ref{xx}) can be 
obtained as follows. Applying expansion $(a+b)^m=\sum _{i=0} ^m 
\binom{m} {i} \ a^{m-i} \ b^i$ and considering Eqs.~(\ref{CosInt}) 
we obtain from eq.~(\ref{rmInak}) the formula
\begin{multline} \label{rBin}
\overline{ \rho_{N+1}^m } = \sum \limits _{i=0} ^{\text{int}(m/2)}
\ \sum \limits _{j=0} ^{m-2i} \, \sum \limits _{n=0} ^{i}
\binom{m}{2i} \binom{m-2i}{j} \binom{i}{n}
\\ \times \overline{\left(\sqrt{\lambda_1^2 -1 } \cos \phi_N
\right)^{2i}} \ \left(\frac{\lambda_1 -1}{2} \right)^j \
\\ \times \lambda_1^{m-2i-j} \ \rho_N^{m-i-j+n}.
\end{multline}
To express eq.~\myref{rBin} in the form~(\ref{xx1}), we choose in
the triple sum of ~\myref{rBin} all terms with $m-i-j+n=k$. We
write all these terms as a single term $\alpha_k(m) \rho_N^k$,
where
\begin{multline} \label{wn}
\alpha_k(m)= \sum \limits _{i=0} ^{\text{int}(m/2)} \ \sum \limits
_{j=0}
^{m-2i} \Theta(k+i+j-m) \ \Theta(m-k-j) \\
\times \binom{m}{2i} \binom{m-2i}{j} \binom{i}{k+i+j-m}
\\ \times \overline{\left(\sqrt{\lambda_1^2 -1 } \cos \phi_N
\right)^{2i}} \ \left(\frac{\lambda_1 -1}{2} \right)^j \
\lambda^{m-2i-j}_1 ,
\end{multline}
with $\Theta(x \ge 0)=1$ and $\Theta(x < 0)=0$. To derive
eq.~(\ref{wn}) we have also regarded the limits $0 \le n \le i$,
which give the conditions $k+i+j-m \ge 0$ and $m-k-j \ge 0$.

For $k=m$ the function $\Theta(m-k-j)$ gives the only solution
$j=0$ and eq.~\myref{wn} reduces to
\begin{equation} \label{wkm2}
\alpha_m(m)= \sum \limits _{i=0} ^{\text{int}(m/2)} \binom{m}{2i}
\lambda^{m-2i}_1 \overline{\left(\sqrt{\lambda_1^2 -1 } \cos
\phi_N \right)^{2i}}.
\end{equation}
Equation~(\ref{wkm2}) is just binomial expansion of eq. \myref{alphm}.

%%%%%%%%%%%%%%%%%%%%%%%%%%%%%%%%%%%%%%%%%%%%%%%
\section{} \label{a2App}

Here we derive the $N$-dependence of $\overline{\rho_N^2}$. In accord
with eq. \myref{gf}, we assume
\begin{equation} \label{r2}
\overline{\rho^2 _N } = a_2(2)\, \lambda_2^N + a_1(2)\, \lambda_1^N +
a_0(2) ,
\end{equation}
where the parameters $\lambda_2$, $a_2(2)$, $a_1(2)$, and $a_0(2)$
have to be determined while $\lambda_1$ is known. Also known are the
coefficients $a_1(1)=1/2$ and $a_0(1)=-1/2$ [compare Eqs.
\myref{xy} and \myref{r1} with \myref{gf} for $m=1$].

Combining Eqs. \myref{rmInak}, \myref{xx1}, and \myref{xx} for m=2
we obtain
\begin{equation} \label{Rho2N}
\overline{\rho _{N+1} ^2} = \alpha_2(2)\, \overline{ \rho^2_N } +
\alpha_1(2)\, \overline{ \rho_N } + \alpha_0(2),
\end{equation}
where $\alpha_2(2)=\lambda_2$, $\alpha_1(2)=\lambda_2-\lambda_1$,
and $\alpha_0(2)=(\lambda_1-1)^2/4$. Inserting Eqs. \myref{r1} and
\myref{r2} into eq. \myref{Rho2N} we get
\begin{multline} \label{s1}
a_2(2)\, \lambda_2 \lambda_2^N + a_1(2)\, \lambda_1 \lambda_1^N +
a_0(2)\,  = \\
\alpha_2(2)\, \left[ a_2(2)\, \lambda_2^N + a_1(2)\, \lambda_1^N +
a_0(2)\, \right] +
\\ \alpha_1(2)\, \left[ a_1(1)\, \lambda_1^N + a_0(1)\, \right] + \alpha_0(2) .
\end{multline}
Now we compare the $N$-independent factors at $\lambda_0^N \equiv
1$, $\lambda_1^N$, and $\lambda_2^N$ on both sides of eq.
\myref{s1}. For $\lambda_0^N$ we obtain
\begin{equation} \label{b0}
a_0(2) = \alpha_2(2)\, a_0(2) + \alpha_1(2)\, a_0(1) + \alpha_0(2),
\end{equation}
where the only unknown parameter is $a_0(2)$. Thus, eq. \myref{b0}
immediately gives
\begin{equation} \label{b0final}
a_0(2) = \frac{1}{3}.
\end{equation}
Analogously, for $\lambda_1^N$ we obtain
\begin{equation} \label{b1}
a_1(2)\, \lambda_1 = \alpha_2(2)\, a_1(2)\, + \alpha_1(2)\, a_1(1),
\end{equation}
so that
\begin{equation} \label{b1final}
a_1(2) = -\frac{1}{2}.
\end{equation}
Eventually, for $\lambda_2^N$ we get
\begin{equation} \label{b2}
a_2(2)\, \lambda_2 = \alpha_2(2)\, a_2(2) ,
\end{equation}
which leads to the already known [see eq. \myref{Rho2N}] identity
$\lambda_2(2) = \alpha_2(2)$. In order to calculate $a_2(2)$ we
have to insert the condition $\overline{\rho_0^m} \equiv 0$ into
\myref{r2}. We get
\begin{equation} \label{Nula}
\overline{\rho^2 _0 } \equiv 0 = a_2(2) + a_1(2) + a_0(2),
\end{equation}
so that
\begin{equation} \label{b2Final}
a_2(2) = \frac{1}{6}.
\end{equation}

%%%%%%%%%%%%%%%%%%%%%%%%%%%%%%%%%%%%%%%%%%%%%%%
%%%%%%%%%%%%%%%%%%%%%%%%%%%%%%%%%%%%%%%%%%%%%%%%%%%%%%%%%%%%%%%%%%%
% Bibliography
%%%%%%%%%%%%%%%%%%%%%%%%%%%%%%%%%%%%%%%%%%%%%%%%%%%%%%%%%%%%%%%%%%%


\begin{thebibliography}{99}
\bibitem{comments}
The ratio of Eqs.~(\ref{tra1}) and~(\ref{XiTyp}) can be written in the form
%
\begin{equation} \label{XiPom}
\frac{\xi}{\xi_1} = \frac{L/\xi_1}{L/\xi} = \frac{\lambda_1-1} {2
\ln\left( \frac{\lambda_1 +1}{2} \right)} , \nonumber
\end{equation}
%
which expresses $\lambda_1$ as an unambiguous function of $L/\xi$ and
$L/\xi_1$. By means of eq.\myref{xim} we can also write equation
%
\begin{equation} \label{LXim}
\frac{L}{\xi_m} =
\frac{L}{\xi_1} \ \frac {2(\lambda_m -1)} {m(m+1) (\lambda_1 -1)}.
\nonumber
\end{equation}
%
Since $\lambda_m$ is an unambiguous function of $\lambda_1$, the
latter equation combined with the former one determines $L/\xi_m$
as an unambiguous function of $L/\xi$ and $L/\xi_1$.
\end{thebibliography}
\end{document}